\documentstyle[aps]{revtex}
\begin{document}
\draft
\tighten
\title{Unification of gauge coupling constants in the minimal
supersymmtric model with $\alpha_s\approx0.11$}

\author{\bf A. K. Chaudhuri\cite{byline}}

\address{ Variable Energy Cyclotron Centre\\
1/AF,Bidhan Nagar, Calcutta - 700 064\\}

\maketitle
\begin{abstract}
We have studied the gauge unification with the recent electroweak
data as a function of the higgsino mass. It was shown that if the
strong  coupling  constant  is  small  $\approx 0.11$, consistent
picture of gauge unification  is  not  possible  in  the  minimal
supersymmetric standard model. \end{abstract}

\pacs{PACS number(s): 12.60.Jv, 12.10.Kt}

\section{Introduction}

One of the testing ground of grand unified theories is the strict
unification of the three coupling constants $\alpha_1$,$\alpha_2$
and  $\alpha_s$  of  the  standard  model $SU(3)_c \times SU(2)_L
\times U(1)_Y$ at some high scale ($M_{GUT}$),  using  as  inputs
their  experimental  values  at  the  Z-pole  mass.  The coupling
constants $\alpha_1$ and $\alpha_2$ can be  determined  from  the
accurate    experimental   measurement   of   $\alpha_{em}$   and
$\sin^2\theta_W$ at $M_z$ pole mass. The world average values for
$M_z$,$\alpha_{em}$ and $\sin^2\theta_W$ are \cite{rev},

\begin{mathletters}
\label{1}
\begin{eqnarray}
M_z=&&91.184 \pm 0.0022\\
\alpha^{-1}_{em}=&&127.9 \pm 0.09\\
\sin^2\theta_W=&&0.2315 \pm 0.0002 \pm 0.0003
\end{eqnarray}
\end{mathletters}

The  coupling constants $\alpha_1$ and $\alpha_2$ of the $U(1)_Y$
and  $SU(2)_L$   gauge   are   related   to   $\alpha_{em}$   and
$\sin^2\theta_W$ as,

\begin{mathletters}
\begin{eqnarray}
\alpha_1=&&\frac{5}{3}\frac{\alpha_{em}}{\cos^2\theta_W}\\
\alpha_2=&&\frac{\alpha_{em}}{\sin^2\theta_W}
\end{eqnarray}
\end{mathletters}

The strong coupling constant ($\alpha_s$) has also been measured,
its  world  average  value  is;  $\alpha_s(M_z)=0.123  \pm 0.005$
\cite{rev}. Recently, it has been pointed out that  QCD  can  not
tolerate  such a large $\alpha_s$ \cite{sh95}. Several low energy
experiments also indicate that $\alpha_s$ must be close  to  0.11
\cite{al93},   $3\sigma$  below  the  Z-peak  value  measured  at
collider  experiments.   Theoretically   clean   deep   inelastic
scattering   experiments   give   $\alpha_s(M_z)=0.112  \pm0.005$
\cite{vi92,ma95}. A new analysis of the $\Upsilon$ sum rule yield
$\alpha_s(M_z)=0.109 \pm 0.001$ \cite{vo95}. Similar  values  are
obtained  in lattice QCD: $\alpha_s(mz)=0.110 \pm 0.006$ ($c\bar{
c}$ spectrum)  \cite{el92}  and  $0.115  \pm  0.002$  ($b\bar{b}$
spectrum)  \cite{da95a}.  The  apparent  conflict between the low
energy and the collider energy Z-peak determination of $\alpha_s$
may be due to the higher order corrections  to  the  LEP  values.
Preliminary  analysis suggests that these higher order correction
may bring down the collider values to be consistent with the  low
energy  measurements  \cite{el91}. Indeed it has also been argued
that the systematic error usually quoted in LEP number is grossly
underestimated, and the LEP  can  only  claim  to  determine  the
strong      coupling      constant      within     the     limit,
$0.10\leq\alpha_s\leq0.15$ \cite{co95}.

It  is  now  established  that the standard model is inconsistent
with gauge unification, with the experimentally measured coupling
constants \cite{am91,el91b,la91}. At early 90,  the  measurements
were  consistent  with  the  gauge  unification  in  the  minimal
supersymmetric  standard  model  (MSSM)   \cite{am91,el91b,la91}.
Recently  Langacker  and  Polonsky  \cite{la95}  using the recent
electroweak data found that in MSSM, $\alpha_s(M_z)\approx 0.129$
is  required  for  strict  gauge  unification.  This   value   is
considerably   higher   than   the   world   average   value   of
$\alpha_s(M_z)=0.123$. Also it  is  much  above  the  low  energy
measurements       and      QCD      motivated      value      of
$\alpha_s(M_z)\approx0.11$. In the  context  of  MSSM  the  lower
$\alpha_s$  require raising the SUSY particle masses considerably
higher  than  the  1  TeV  scale.  However,  SUSY  mass   spectra
considerably  higher  than  1  TeV  scale  will  have  problem of
diverging radiative correction. Also, heavy SUSY masses  will  be
inconsistent    with    our   expectation   that   the   lightest
supersymmetric particle (LSP) will be neutral and  the  candidate
for   the  dark  matter.  The  other  possibility,  to  reconcile
$\alpha_s\approx 0.11$ with gauge unification, is to assume  very
large  negative heavy threshold correction at GUT scale and NRO's
which     could     decrease     $\alpha_s(M_z)$     by      10\%
\cite{la95,ri95,da95b}.    Recently,   Roszkowski   and   Shifman
\cite{ro96} argued that in the simple supersymmetric extension of
the standard model, one need not have the constraint of universal
gaugino masses at the GUT scale, which is, in general assumed  in
MSSM.  To  fully  specify  the  MSSM,  one  needs  to  invoke the
supersymmetry breaking pattern. The soft breaking term, which  in
essence  couple  the  MSSM with the N=1 supergravity give rise to
the constraint of universal gaugino masses at GUT scale. Thus  in
a  pure phenomenological approach, one can relax the condition of
universal gaugino  masses.  They  showed  that  by  relaxing  the
constraint   small   $\alpha_s$   compatible   with   low  energy
measurements can be obtained\cite{ro96}.  In  that  case,  gluino
becomes  lighter  than  the  wino  and  can be well below 200 GeV
\cite{ro96}.  However, the approch is unsatisfactory, as there is
no theoretical scheme for the symmetry breaking term.

In  the  present  paper, we show that consitent gauge unification
with small $\alpha_s$ is not possible  within  the  framework  of
MSSM.  Our approah is phenomenological. We keep the constraint of
universal gaugino masses. We  treat  the  higgsino  masses  as  a
parameter of unification, and tune it to obtain unification for a
given  value  of  $\alpha_s$.  Gauge  unification  with $\alpha_s
\approx$ 0.11, require higgsino masses in  the  range  of  $10^6$
GeV,  three  order  of  maginitude  larger  than  other sparticle
masses. Higgsino masses in the range of $10^6$ GeV  will  require
that  the  supersymmetry  break  at that scale only and the model
will face the unsatisfactory aspect of gauge hierarchy problem.

The  paper is organised as follows. In section 2, we describe the
model briefly,  in  section  3,  the  results  obtained  will  be
discussed. Summary and conclusions will be given in section 4.

\section{ Minimal supersymmetric standard model (MSSM)}

The renormalisation group equations for the gauge  couplings  are
given by (neglecting the small Yukawa couplings),

\begin{equation}
\frac{d\alpha^{-1}_i}{dt}=-\frac{b_i}{2\pi}
-\sum_j\frac{b_{ij}\alpha_j}{8\pi^2}, i=1,3
\end{equation}

\noindent  where $t=\ln(Q/M_{GUT})$, with Q the running scale and
$M_{GUT}$ the unification scale mass.

The  one loop coefficients $b_i$ of the $\beta$ functions for the
gauge couplings change across each new running mass threshold. In
the MSSM they can be parameterized as \cite{ro92,el92b,ka94},

\begin{mathletters}
\begin{eqnarray}
b_1         =         &&        \frac{41}{10}+        \frac{2}{5}
\theta_{\tilde{H}}+\frac{1}{10}\theta_{H_2}
+\frac{1}{5}\sum^3_{i=1} [\frac{1}{12}(\theta_{\tilde{u}_{L_i}}+
             \theta_{\tilde{d}_{L_i}})
+\frac{4}{3} \theta_{\tilde{u}_{R_i}}
+\frac{1}{3} \theta_{\tilde{d}_{R_i}}
+\frac{1}{4}(\theta_{\tilde{e}_{L_i}}
+            \theta_{\tilde{\nu}_{L_i}})
            +\theta_{\tilde{e}_{R_i}}]\\
b_2  = && -\frac{19}{6}+ \frac{4}{3} \theta_{\tilde{W}}
+ \frac{2}{3}\theta_{\tilde{H}}
+\frac{1}{6}\theta_{H_2}
+\frac{1}{2}\sum^3_{i=1} [\theta_{\tilde{u}_{L_i}}
\theta_{\tilde{d}_{L_i}} +
\frac{1}{3} \theta_{\tilde{e}_{L_i}}
\theta_{\tilde{\nu}_{L_i}}]\\
b_3   =   &&   -7+  2\theta_{\tilde{g}}+  \frac{1}{6}\sum^3_{i=1}
[\theta_{\tilde{u}_{L_i}}+       \theta_{\tilde{d}_{L_i}}       +
\theta_{\tilde{u}_{R_i}}+ \theta_{\tilde{d}_{R_i}}]
\end{eqnarray}
\end{mathletters}

\noindent where $\theta_x \equiv \theta(Q^2-m^2_x)$. In the above
equations  $\tilde{H}$  stands  for  the mass degenerate Higgsino
fields, $\tilde{W}$ for the winos, the partner of the SU(2) gauge
bosons ($M_{\tilde{W}}\equiv M_2$), $\tilde{g}$  is  the  gluino,
the partner of the gluon, all are assumed to be mass eigen states
in this approximation. The $H_2$ stands for the heavy Higgs boson
doublet \cite{ka94}.

The  effect  of  low  mass threshold on two loop beta function is
expected to be small. In the weak  scale,  we  thus  use  the  SM
values:

\begin{equation}
b_{ij}=\left   (\matrix{   \frac{199}{50}   &   \frac{27}{10}   &
\frac{44}{5}\cr \frac{9}{10} & \frac{35}{6} & 12\cr
               \frac{11}{10} & \frac{9}{2} & -26\cr}\right )
\end{equation}

\noindent and from the weak to the GUT scale, we use,

\begin{equation}
b_{ij}=\left    (\matrix{   \frac{199}{25}   &   \frac{27}{5}   &
\frac{88}{5}\cr \frac{9}{5} & 25 & 24\cr
               \frac{11}{5} & 9 & -14\cr}\right )
\end{equation}

The   RGE   equations  can  be  integrated  in  a  step  function
approximation to obtain ${\alpha \prime}_{i}(\mu  \prime)$  at  a
scale $\mu \prime$ for a given $\alpha_i(\mu)$,

\begin{equation}
\frac{1}{\alpha \prime_i(\mu \prime)}=\frac{1}{\alpha_i(\mu)}+
\beta_0 \ln\frac{\mu \prime}{\mu}
+\frac{\beta_1}{\beta_0}\ln \frac{1/\alpha \prime_i(\mu \prime)
+\beta_1/\beta_0}
{1/\alpha_i(\mu)+\beta_1/\beta_0}
\label{2}
\end{equation}

\noindent with

\begin{mathletters}
\begin{eqnarray}
\beta_0=&&-\frac{1}{2\pi}(b_i+\frac{b_{ij}}{4\pi}\alpha_j(\mu)+
\frac{b_{ik}}{4\pi}\alpha_k(\mu))\\
\beta_1=&&- \frac{2b_{ii}}{(4\pi)^2}
\end{eqnarray}
\end{mathletters}

The  eq.\ref{2}  can be solved iteratively to obtain the coupling
constants at any arbitrary energy, knowing their value at a given
energy.

In  addition to the above equations, we consider the evolution of
gaugino masses. There evolution equations are simple,

\begin{equation}
\frac{dM_i}{dt}=-\frac{b_i}{4\pi} \alpha_i M_i
\end{equation}

\noindent    with    the   boundary   condition   at   $M_{GUT}$:
$M_i(t=0)=M_{1/2}$. The solution for the gluino and the winos can
be written as,

\begin{mathletters}
\begin{eqnarray}
M_{\tilde{W}}=&&\frac{\alpha_2(M_{\tilde{W}})}{\alpha_5}
M_{1/2}\\
M_{\tilde{g}}=&&\frac{\alpha_s(M_{\tilde{g}})}{\alpha_5}
M_{1/2}
\end{eqnarray}
\end{mathletters}

\noindent  where  $\alpha_5$  is the unified coupling constant at
the GUT scale. Combining the two equation we obtain,

\begin{equation}
M_{\tilde{g}}
=\frac{\alpha_s(M_{\tilde{g}})}{\alpha_2(M_{\tilde{W}})}
M_{\tilde{W}} \label{3}
\end{equation}

For  a  given  $M_{\tilde{W}}$,  $M_{\tilde{g}}$  can  be  obtain
iteratively from eq.\ref{3}

The  simple  step-function approximation used to integrate the RG
equations is justified only in $\overline{DR}$  scheme.  However,
the experimental $\alpha_{em}$ and $\sin^2\theta_W$ were obtained
in  the $\overline{MS}$ scheme. We therefore convert the coupling
constants into the $\overline{DR}$ scheme by,

\begin{equation}
\frac{1}{\alpha_{i}^{\overline{DR}}}
=\frac{1}{\alpha_{i}^{\overline{MS}}}
-\frac{C_{i}}{12\pi}
\end{equation}

\noindent where $C_1=0,C_2=2$ and $C_3=3$.

\section{Results}

\subsection{Higgsinos and gauge unification}

We  note  that  apart  from  the  winos  and the gluino, the beta
function coefficients are most affected  by  the  higgsinos.  Not
only  they  appear in two beta function $b_1$ and $b_2$, compared
to other sparticles,  their  coefficients  are  also  larger.  We
therefore choose the higgsino mass as the parameter which will be
tuned  to  obtain strict gauge unification with recent electroweak
data. The wino and the gluino will be treated separately as  they
are connected by the universal gaugino mass at GUT scale. All the
other  SUSY  masses  will be assumed to be degenerate at a common
mass ($M_c$). This is certainly an assumption, however, it  helps
to identify the role of higgsinos in the unification process.

We  define  the  GUT scale $(M_{GUT})$ as the scale where all the
three coupling constants unify at some value $\alpha_5$.  The  RG
equations  for  $\alpha_1$,  $\alpha_2$  and  $\alpha_3$ were run
simultaneously from $M_z$ to $M_{GUT}$. The input $\alpha_1$  and
$\alpha_2$  were  calculated  from the experimental $\alpha_{em}$
and $\sin^2\theta_W$ (eq.\ref{1}). We have used the central value
for $\alpha_{em}$. Unification of three forces  is  sensitive  to
the  input  value of $\sin^2\theta_W$. This sensitivity is due to
the fact that $\alpha_2$ does not change much between  $M_z$  and
$M_{GUT}$,  as do the other two couplings. Thus a small change in
$\sin^2\theta_W$ has an enhanced effect on  the  unification.  We
therefore   consider   two   input  values  of  $\sin^2\theta_W$:
$\sin^2\theta_W=0.231$ and $\sin^2\theta_W=0.232$ , covering  the
$1\sigma$  variance. For the input strong coupling $\alpha_s$, we
choose  a  value  between  0.10-0.13.  For  a  given  wino   mass
($M_{\tilde{W}}$)  and  the  common SUSY mass $M_{c}$ we vary the
higgsino mass ($M_{\tilde{H}}$)  100  GeV  onwards  to  find  the
minimum  $M_{\tilde{H}}$  required  for the strict unification of
the three couplings. The gluino  mass  was  obtained  iteratively
from eq.\ref{3}.

In  fig.1a,  we  have  shown  the  higgsino  mass $M_{\tilde{H}}$
required to obtain strict gauge unification, as a function of the
input $\alpha_s$. The common SUSY particle  mass($M_c$)  and  the
wino  mass  were  fixed  at 1000 GeV. The black dots and the open
triangles   corresponds   to    $\sin^2\theta_W=    0.231$    and
$\sin^2\theta_W=  0.232$ respectively. The higgsino mass required
for  unification  shows  a  sensitive  dependence  on  the  input
$\alpha_s$.  It  also  depends on the input $\sin^2\theta_W$. For
$\sin^2\theta_W$=0.231,  the   higgsino   mass   varies   between
$6.4\times10^9$  GeV to $4.5\times10^3$GeV, as $\alpha_s$ changes
from 0.10 to 0.13. The higgsino mass is lowered approximately  by
a factor of 10, if $\sin^2\theta_W=0.232$. It then varies between
$8.2\times10^8$  GeV  to  $6\times10^2$GeV.  In fig.1b and 1c, we
have shown the unification scale ($M_{GUT}$) and the  inverse  of
the  unified  coupling  ($\alpha^{-1}_5$).  We note that they are
anti-correlated.  While  $M_{GUT}$   increases,   $\alpha^{-1}_5$
decreases  with  the  input  $\alpha_s$.  In  comparison  to  the
higgsino mass,  $M_{GUT}$  is  quite  insensitive  to  the  input
$\alpha_s$.  As  $\alpha_s$  changes  from  0.10 to 0.13, the GUT
scale  is  increased  by  a  factor   of   2   only.   Similarly,
$\alpha^{-1}_5$   also   shows   weak  dependence  on  the  input
$\alpha_s$. It is changed by less than 10\%.  Dependence  of  the
GUT  scale on the input $\sin^2\theta_W$ is also manifest. Within
$1\sigma$  variation   of   $\sin^2\theta_W$,   it   is   changed
approximately by 10\%. Interestingly, the unified coupling do not
show  appreciable  dependence  on  the input $\sin^2\theta_W$. In
fig.1d, the ratio $M_{\tilde{g}}/M_{\tilde{W}}$ is shown.  Within
$1\sigma$  variation  of $\sin^2\theta_W$ the gluino mass also do
not show  appreciable  dependence  on  the  input  value  of  the
Weinberg  angle. However, as input $\alpha_s$ varies from 0.10 to
0.13, it increases from  2300  GeV  to  2800  GeV,  showing  weak
dependence  on  the input $\alpha_s$. In fig.1e, we have depicted
the variation of the universal gaugino mass $M_{1/2}$.  $M_{1/2}$
also   do   not   show   appreciable   dependence  on  the  input
$\sin^2\theta_W$.  It  shows  weak  dependence   on   the   input
$\alpha_s$,  increasing  from $1.15\times10^3$ GeV to $1.23\times
10^3$ GeV, as $\alpha_s$ changes from 0.10 to 0.13.

The  results  discussed  so  far indicate that while the higgsino
mass shows a sensitive dependence to  the  input  $\alpha_s$  for
strict  unification,  (it  changes  by  6  order  of magnitude as
$\alpha_s$ changes from 0.10 to 0.13)  other  parameters  of  the
model,  e.g. $M_{GUT}$, $\alpha_5$, $M_{\tilde{g}}$ and $M_{1/2}$
shows a weak dependence on the input $\alpha_s$. Also, unlike the
higgsino mass, the input dependence of $\sin^2\theta_W$  is  also
less  in  those  parameters.  With  the  world  average  value of
$\alpha_s$=0.123, strict unification is  possible  with  Higgsino
mass  as  well  as all the other sparticle masses in the range of
TeV scale. However,  if  we  consider  the  QCD  motivated  small
$\alpha_s  \approx$0.11,  strict  unification  is  obtained  with
Higgsino masses in the range of $10^6-10^7$ GeV.  All  the  other
SUSY  masses  can  be  in  the  TeV  scale.  To  be specific, for
$\alpha_s=0.11$,  strict  gauge  unification  is  obtained   with
$M_{\tilde{H}}=1.46\times10^7     (2.15\times10^6$)    GeV    for
$\sin^2\theta_W$=0.231(0.232).

In    the   above   calculations,   the   mass   of   the   winos
($M_{\tilde{W}}$) were fixed at 1000 GeV. To observe  the  effect
of  wino  mass  on the unification, with small $\alpha_s$, we now
vary $M_{\tilde{W}}$ from 100 GeV to 2000 GeV. We fix the  strong
coupling at $\alpha_s=0.11$. The common SUSY mass $M_c$ was fixed
at  1000  GeV.  As  before,  while the central value was used for
$\alpha_{em}$, for $\sin^2\theta_W$, we consider variation within
$1\sigma$. In fig.2,  the  results  are  shown.  In  fig.2a,  the
higgsino  mass  required  for  strict  unification  is shown as a
function of the wino mass. As observed earlier, the higgsino mass
depends on  the  input  $\sin^2\theta_W$.  Within  its  $1\sigma$
variation,  it  is  changed  approximately  by  a  factor  of 10.
However, the higgsino mass required for unification shows a  weak
dependence  on the wino mass. As the wino mass is varied from 100
to 2000 GeV, the higgsino mass is decreased  by  a  factor  of  5
only.  Interestingly,  we  find  that  after  1000  GeV,  the the
higgsino mass remains nearly same. Variation of the GUT scale and
inverse of the unified coupling with the wino mass  is  shown  in
fig.2b  and  2c.  As  before,  they  are  anti-correlated.  While
$M_{GUT}$ shows weak dependence on  the  input  $\sin^2\theta_W$,
$\alpha^{-1}_5$   do  not  exhibit  any  such  dependence.  There
dependence on the wino  mass  is  also  weak.  Here  also,  after
$M_{\tilde{W}}$=1000  GeV,  rate  of  variation  of $M_{GUT}$ and
$\alpha^{-1}_5$  slows  down.  The   variation   of   the   ratio
$M_{\tilde{g}}/M_{\tilde{W}}$  with  the  wino  mass  is shown in
fig.2d. The ratio do not depend on the input $\sin^2\theta_W$. It
decreases with the wino mass. Again after 1000 GeV, the  rate  of
decrease  is  slowed  down and the relation $M_{\tilde{g}}\approx
2.5M_{\tilde{W}}$ become valid. In fig.2e, the universal  gaugino
mass  $M_{1/2}$  and  its  variation with the wino mass is shown.
$M_{1/2}$   do   not   show   any   dependence   on   the   input
$\sin^2\theta_W$.  As  expected,  it  increases linearly with the
wino mass. The results indicate that  the  unification  of  three
forces  with  $\alpha_s=0.11$  depend  weakly upon the wino mass.
Increasing the wino mass beyond 1000 GeV,  does  not  affect  the
unification appreciably.

We have also studied the dependence of the common SUSY mass $M_c$
on  the  unification  process with $\alpha_s$=0.11. The wino mass
was fixed at 1000 GeV. In fig.3a, the variation of  the  higgsino
mass   with   $M_c$   is   shown.   The   input   dependence   of
$\sin^2\theta_W$ on the higgsino  mass  is  again  manifest.  The
higgsino  mass  shows  a  very weak dependence on the common SUSY
mass $M_c$.  Similarly,  the  GUT  scale  (fig.3b),  the  unified
coupling   (fig.3c),   the   ratio  $M_{\tilde{g}}/M_{\tilde{W}}$
(fig.3d) and $M_{1/2}$ (fig.3e) shows very weak dependence on the
input  $M_c$.  Thus  the  unification  of   three   forces   with
$\alpha_s=0.11$  is  insensitive  to  the  common SUSY mass $M_c$
within the range 500-2500 GeV. Thus a  more  accurate  SUSY  mass
spectra will not alter the present results significantly.

The  present  analysis  indicate  that  low  energy QCD motivated
strong coupling constant $alpha_s\approx 0.11$ is consistent with
gauge unification in MSSM, if the  higgsino  masses  are  in  the
range  of  $10^6-10^7$  GeV  and other SUSY masses are in the TeV
range. In the following we will examine its effect on some  other
features of the model.

\subsection{Color triplet Higgs mass}

Any GUT theory must consider the nucleon decay rate in the model.
As  the GUT scale obtained presently exceeds $10^{15}$ GeV, we do
not expect problem from the  direct  proton  decay  $p\rightarrow
e^+\pi^0$.   However,   nucleons   can   decay  via  exchange  of
color-triplet  Higgs  multiplet  \cite{sa82,we82},  through   the
dimension   five   operator.   The   color   triplet  Higgs  mass
$M_{H_{c}}$, then has significant observable  effect.  The  lower
bound  on  $M_{H_{c}}$  from  nucleon-decay  experiments  can  be
obtained as \cite{hi93},

\begin{equation}
M_{H_{c}}    \geq   5.3\times   10^{15}   GeV
\end{equation}

We  have  calculated  the  color triplet Higgs mass in our model.
Following \cite{hi93,ca96}, we write at the GUT scale,

\begin{equation}
(3\alpha^{-1}_2 -2\alpha^{-1}_3 -\alpha^{-1}_1)(M_{GUT})=
\frac{1}{2\pi}(\frac{12}{5} \ln \frac{M_{H_{c}}}{M_{GUT}}-2
\ln \frac{M_{SUSY}}{\Lambda})
\end{equation}

The  first  term  of  the  rhs  corresponds  to  heavy  threshold
correction at GUT scale as a function of the color triplet  higgs
mass  $M_{H_{c}}$  \cite{hi93,ca96}. The second term is the light
threshold correction at the weak scale ($\Lambda$). (We note that
in the calculations presented at earlier sections,  we  have  not
taken  into  account the heavy threshold effects. However, if the
heavy thresholds are much heavier than the GUT  scale,  then  the
results  obtained  previously  will  not  be altered). For strict
unification at the GUT scale, the lhs of the  above  equation  is
zero and we can find the color triplet Higgs mass in terms of the
SUSY  masses.  In  the  above equation all the SUSY particles are
assumed to be degenerate at a common mass $M_{SUSY}$. However, as
we have treated the winos, gluino and the  higgsinos  separately,
the the second term is to be replaced by

\begin{equation}
2\ln           \frac{M_{SUSY}}{\Lambda}\rightarrow 4          \ln
\frac{M_{\tilde{W}}}{M_{\tilde{g}}}               +\frac{2}{5}\ln
\frac{{M_{c}}}{\Lambda}+\frac{8}{5}                           \ln
\frac{M_{\tilde{H}}}{\Lambda}
\end{equation}

In  fig.4, we have shown the variation of the color triplet Higgs
mass $M_{H_{c}}$ with the wino mass $M_{\tilde{W}}$  as  obtained
from  strict  gauge unification. The strong coupling was fixed at
0.11. The common SUSY mass  $M_c$  was  fixed  at  1000  GeV.  As
discussed  earlier,  for  a  given  wino mass the unification was
obtained by varying the higgsino mass only. Color  triplet  higgs
mass  depend  on  the  input  value  of  $\sin^2\theta_W$,  lower
$\sin^2\theta_W$ results in a higher value for the color  triplet
Higgs  mass. Within $1\sigma$, the value is changed approximately
by a factor of 10. $M_{H_{c}}$ also depend on the wino  mass.  As
the  wino mass is increased its value is decreased. However, 1000
GeV onwards, the rate of decrease is slowed down. We observe that
the color triplet Higgs mass $M_{H_{c}}$ as obtained presently is
much above the lower limit ($5.3\times10^{15}$ GeV)  required  to
respect  the  experimental nucleon decay rate. The result suggest
that large higgsino  mass  will  not  cause  rapid  proton  decay
through the dimension five operator.

\subsection{LSP and dark matter}

About  the  LSP  being the candidate dark matter, we observe that
the higgsinos mixes with gauginos (both are spin 1/2 particle) to
give the neutralino's. In the limiting case when $M_1$,$M_2$  and
$M_{\tilde{H}}$  >>  $M_z$,  the bino and the neutral wino do not
mix with each other nor they mix with the higgsinos.  Thus  binos
can  still be the LSP and the candidate for the dark matter. Thus
if only the higgsinos are in the mass range $10^6-10^7$ GeV,  and
all  the  other  s-particles are in the TeV range, the model will
not suffer from radiative correction at  the  electroweak  scale.
Also  the  expectation  that bino's are the dark matter candidate
need not be altered.

\subsection{Gauge hierarchy problem}

One  of  the  motivation  of introducing the supersymmetry is the
gauge hierarchy problem, i.e. existence of very small  and  large
mass  scale  in  nature.  In SU(5), as well known, the tree level
parameters needed to be fine tuned to an accuracy  of  $10^{-26}$
or  so,  in  order  to  obtain  the  mass  ratio  $M_x/m_W\approx
10^{12}$. However, tree level fine tunings are  upset  at  higher
order  due  to  quadratic  radiative  corrections and SU(5) is in
problem. This need not happen in supersymmetric theories  due  to
the  nonrenormalisation  theorem  of  Grisaru,  Rocek  and Siegel
\cite{gr79}. According to this theory, which is valid  for  exact
or  softly  broken  supersymmetry,  the  parameters  of the super
potential do not only receive infinite renormalisation  but  they
also do not receive finite renormalisation in higher orders. Thus
the   parameters,  once  fine  tuned  at  tree  level  to  obtain
hierarchy, the radiative correction donot disturb  the  hierarchy
at   higher   order.   Thus   problem   of   gauge  hierarchy  in
supersymmetric model is not as  severe  as  in  SU(5).  Also,  in
supersymmetric  theories,  logarithmic  dependence  of  radiative
corrections makes the emergence of high mass  scale  from  a  low
mass scale quite natural.

However,  with  the  Higgsino  mass  in  the scale of $10^6$ GeV,
effective breaking of supersymmetry  will  occur  at  that  scale
only. The effective theory below that scale will not be protected
by  the non-renormalisation theory. Quadratic divergences will be
generated in Higgs mass due to absence of  Higgs-Higgsino-gaugino
coupling at low energy. With Higgsino mass in the range of $10^6$
GeV,  the corrections to the Higgs mass will be of that scale and
correct electroweak  scale  can  not  be  achieved  without  fine
tuning.  Thus  we  find  that  if the strong coupling constant is
small as obtained in  QCD  experiments,  gauge  unification  with
recent  electroweak  data  can  be  achieved, alongwith universal
gaugino masses, only if we abandon one of the primary  motivation
of  supersymmetry,  namely  the resolution to the gauge hierarchy
problem. Also generating  three  order  of  hierarchy  among  the
sparticles  will  be  difficult  to  obtain  in  any  theoretical
framework.  The  discussion  suggests   that   consistent   gauge
unification  with  $\alpha_s \approx 0.11$ is not possible within
the frame work of MSSM.

\section{Summary and conclusions}

We have studied the gauge unification in the MSSM with respect to
the higgsino masses. The purpose of our study was to find whether
with $\alpha_s \approx 0.11$ gauge unification is possible in the
MSSM  with  universal  gaugino mass at GUT scale. Our approach is
phenomenological. We choose the higgsino mass as a parameter. For
a given wino mass, it was tuned to obtain gauge unification.  The
gluino  mass  was obtained iteratively from the condition that at
GUT scale the winos and the gluino have a common mass  $M_{1/2}$.
All the other sparticle masses were assumed to be degenerate at a
common  mass $M_{c}$. The RG equations were run from $M_z$ to the
GUT scale,  with  the  beta  function  coefficients  changing  at
appropriate  masses.  We  find  that if $\alpha_s$ ranges between
0.10 to 0.13, with the recent electroweak data, it is possible to
unify all the three forces by varying the higgsino  mass  between
$10^9-10^3$  GeV.  All  the  other  sparticles  can be in the TeV
scale. The result is insensitive to the wino mass or  the  common
mass  $M_c$. Thus a more accurate SUSY spectra will not alter the
results significantly. It was found  that  if  $\alpha_s  \approx
0.11$,  as measured in QCD experiments, gauge unification require
higgsino masses in the  range  of  $10^6$  GeV,  three  order  of
magnitude  higher  than  other  sparticle  masses. Colour triplet
Higgs were also calculated and found to be heavy enough to forbid
rapid proton decay. However, with Higgsino masses in the scale of
$10^6$ GeV, supersymmetry breaks at that scale only  and  problem
of  gauge  hierarchy resurface. One of the fundamental motivation
of introducing supersymmetry is then lost.

To  conclude,  if  $\alpha_s$ is indeed small, as measured in QCD
experiments, then consistent gauge unification within MSSM is not
possible. However, MSSM  is  merely  the  simplest  extension  of
standard  model  with  supersymmetry.  It is possible that with a
more complicated model a consistent picture will emerge.

\begin{figure}
\caption{Variation  of  (a)  the higgsino mass ($M_{\tilde{H}}$),
(b) the unification scale  ($M_{GUT}$),(c)  the  inverse  of  the
unified       coupling       ($\alpha^{-1}_5$),(d)the       ratio
$M_{\tilde{g}}/M_{\tilde{W}}$ and (e) the universal gaugino  mass
($M_{1/2}$),  with  the strong coupling $\alpha_s(M_z)$ is shown.
The wino mass $M_{\tilde{W}}$ and the common SUSY mass $M_c$  was
fixed  at 1000 GeV. The black dots are for $\sin^2\theta_W=0.231$
and  the  open   triangles   are   for   $\sin^2\theta_W$=0.232.}
\end{figure}
\begin{figure}
\caption{Variation  of  (a)  the higgsino mass ($M_{\tilde{H}}$),
(b) the unification scale  ($M_{GUT}$),(c)  the  inverse  of  the
unified       coupling       ($\alpha^{-1}_5$),(d)the       ratio
$M_{\tilde{g}}/M_{\tilde{H}}$ and (e) the universal gaugino  mass
($M_{1/2}$),  with  the the wino mass ($M_{\tilde{W}}$) is shown.
The strong coupling was fixed at $\alpha_s(M_z)$=0.11. The common
SUSY mass $M_c$ was fixed at 1000 GeV. The  black  dots  are  for
$\sin^2\theta_W=0.231$   and   the   open   triangles   are   for
$\sin^2\theta_W$=0.232.}
\end{figure}
\begin{figure}
\caption{Variation  of  (a)  the higgsino mass ($M_{\tilde{H}}$),
(b) the unification scale  ($M_{GUT}$),(c)  the  inverse  of  the
unified       coupling       ($\alpha^{-1}_5$),(d)the       ratio
$M_{\tilde{g}}/M_{\tilde{H}}$ and (e) the universal gaugino  mass
($M_{1/2}$),  with the the common SUSY mass ($M_c$) is shown. The
strong coupling was fixed at $\alpha_s(M_z)$=0.11. The wino  mass
$M_c$   was   fixed   at   1000  GeV.  The  black  dots  are  for
$\sin^2\theta_W=0.231$   and   the   open   triangles   are   for
$\sin^2\theta_W$=0.232.}
\end{figure}
\begin{figure}
\caption{Variation of the color triplet Higgs mass ($M_{H_{c}}$),
with  the  the  wino  mass ($M_{\tilde{W}}$) is shown. The strong
coupling was fixed at $\alpha_s(M_z)$=0.11. The common SUSY  mass
$M_c$   was   fixed   at   1000  GeV.  The  black  dots  are  for
$\sin^2\theta_W=0.231$   and   the   open   triangles   are   for
$\sin^2\theta_W$=0.232.}
\end{figure}

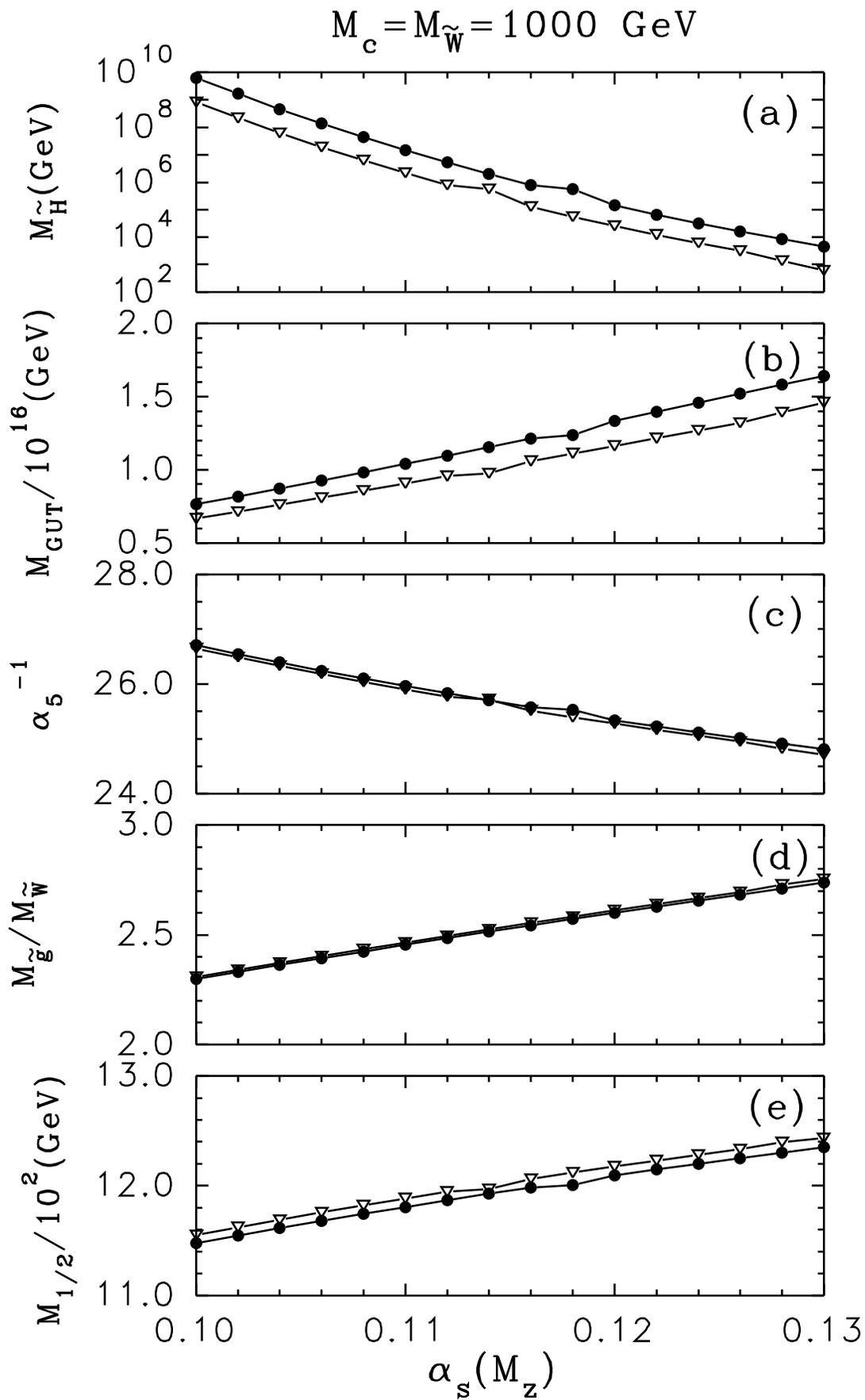
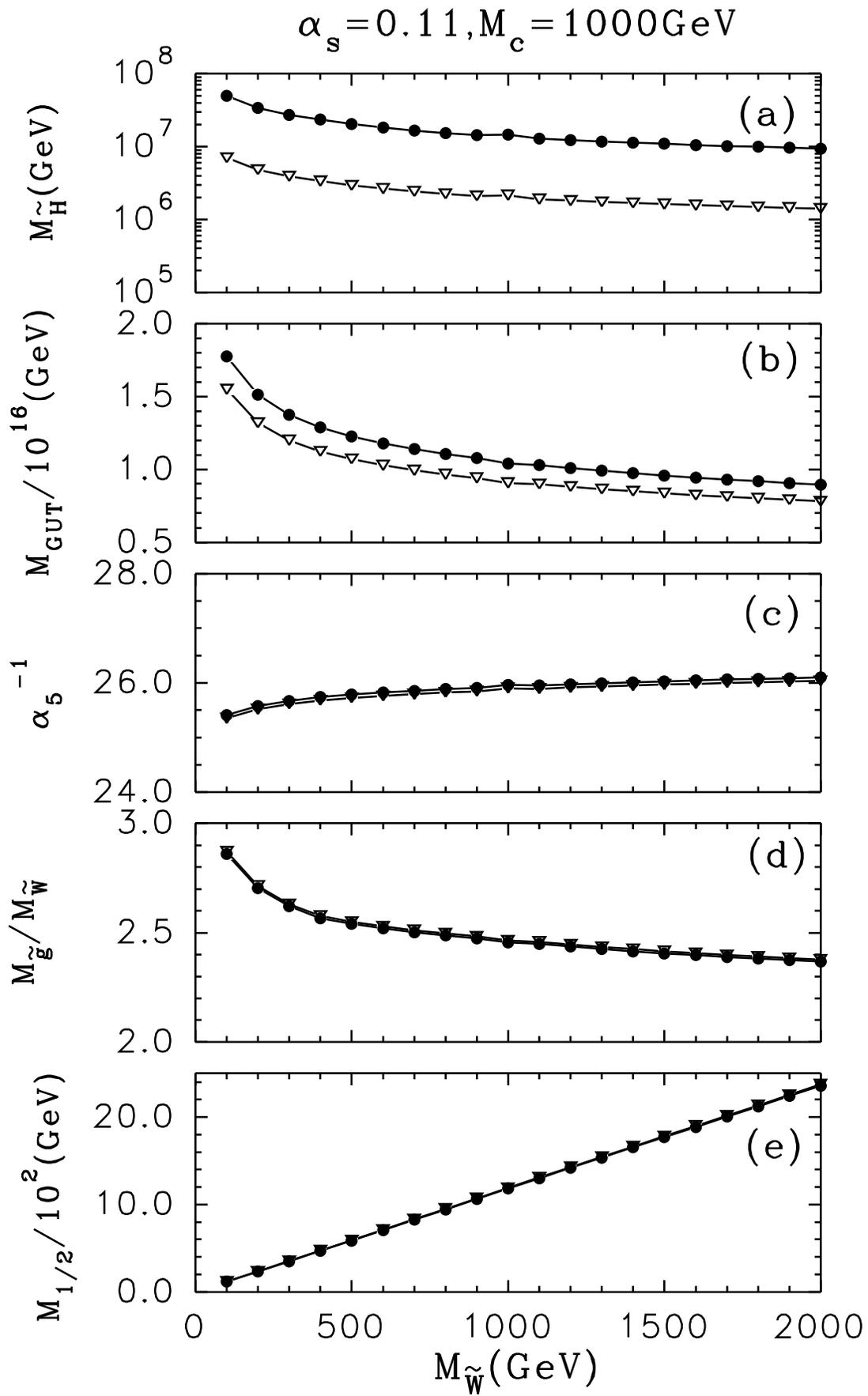
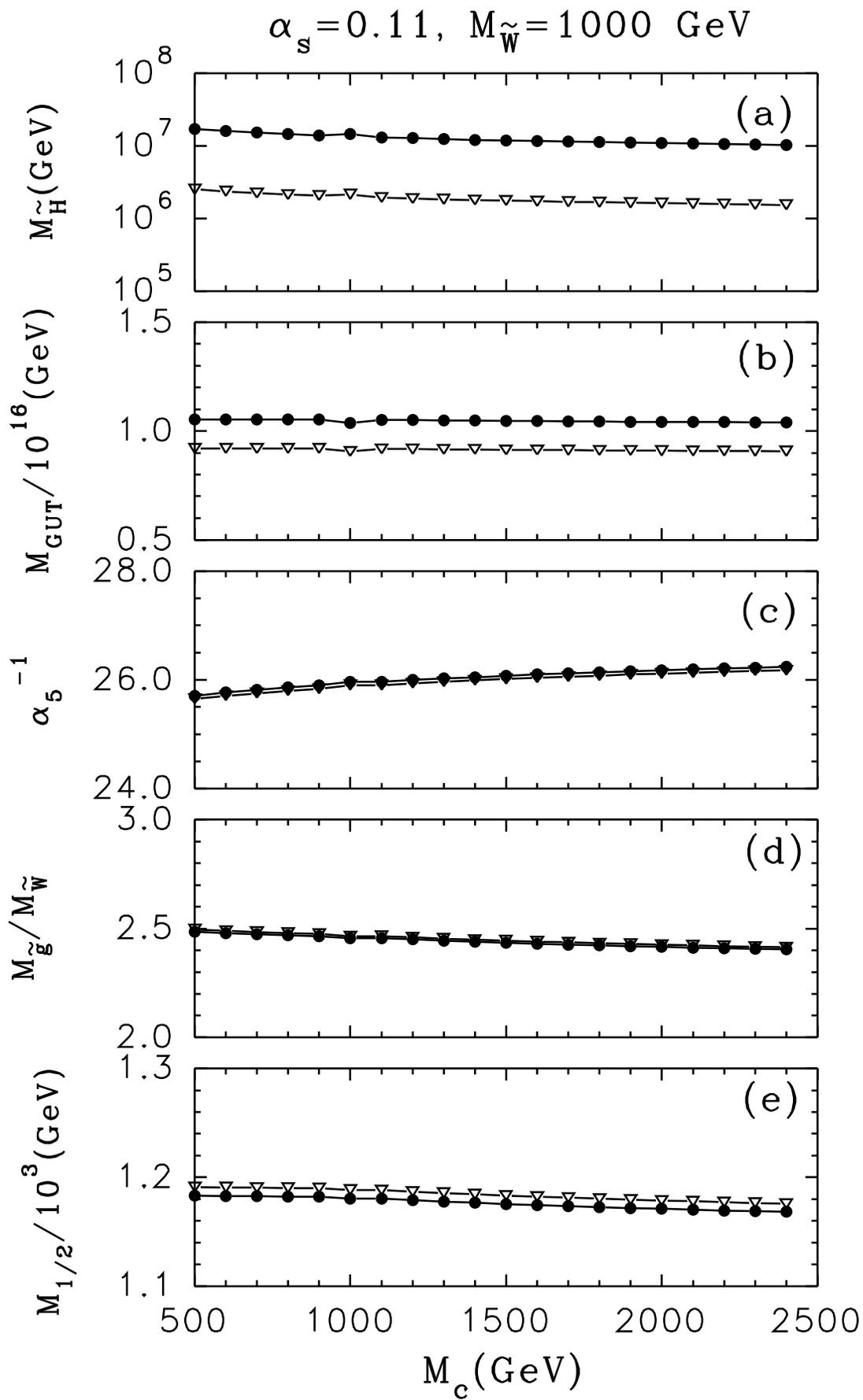
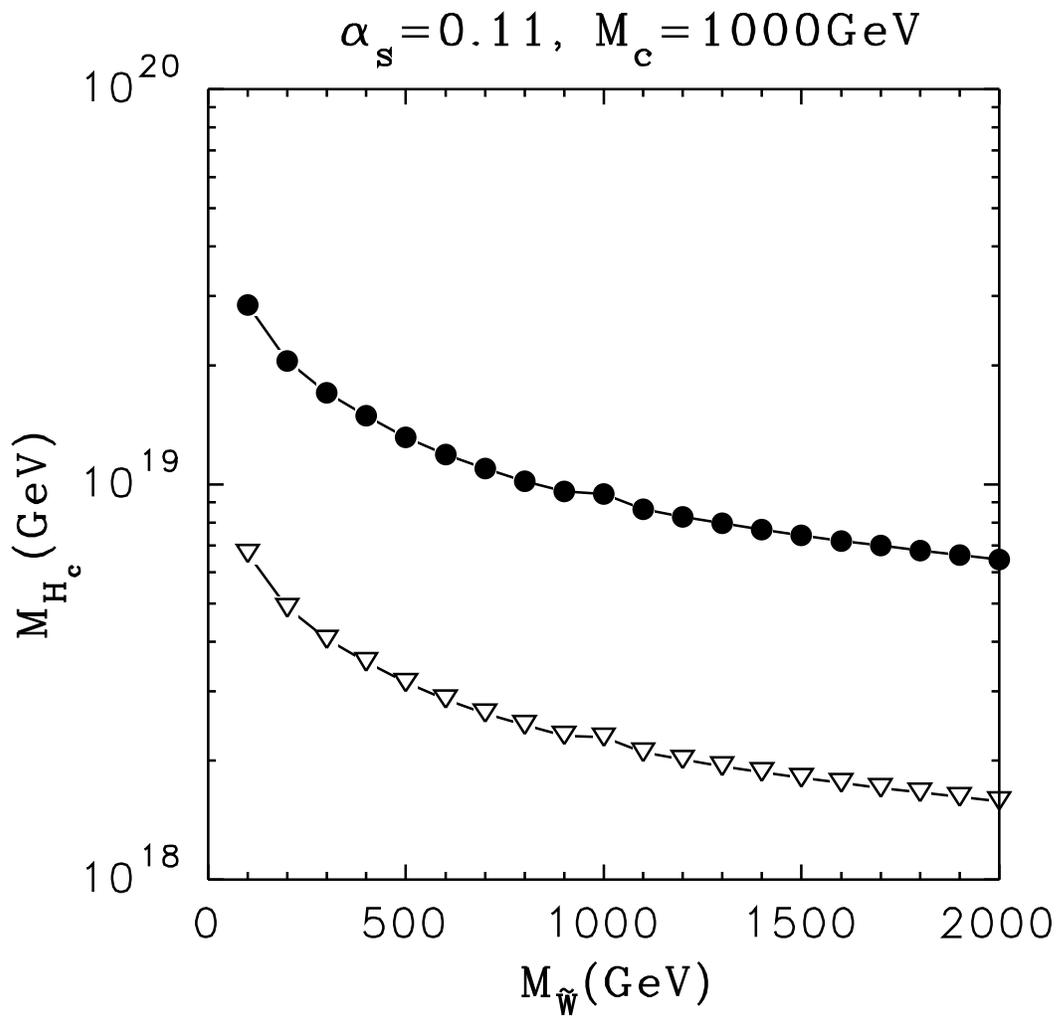
\end{document}